\documentclass[aps,prb,twocolumn,amsmath,amssymb,superscriptaddress,floatfix]{revtex4-2}
\pdfoutput=1

\usepackage{mathrsfs}
\usepackage{amsmath}
\usepackage{amsfonts}
\usepackage{float}
\usepackage{graphicx} 
\usepackage[caption=false]{subfig}
\usepackage{bm} 
\usepackage{xcolor}
\usepackage{epstopdf}
\usepackage{comment}

\newcommand{\kp}{k^\prime}
\newcommand{\gk}{\gamma_k}
\newcommand{\gkd}{\gamma_k^\dagger}
\newcommand{\gnk}{\gamma_{-k}}
\newcommand{\gnkd}{\gamma_{-k}^\dagger}

\newcommand{\ket}[1]{\left | #1 \right\rangle}
\newcommand{\bra}[1]{\left\langle #1 \right|}
\newcommand{\avg}[1]{\left\langle #1 \right\rangle}

\begin{document}
\title{Loschmidt amplitude spectrum in dynamical quantum phase transitions}

\affiliation{ \textit Department of Physics, City University of Hong Kong, Hong Kong}

\author{ Cheuk Yiu Wong }
\affiliation{ \textit Department of Physics, City University of Hong Kong, Hong Kong}
\author{ Wing Chi Yu }
\email{wingcyu@cityu.edu.hk}
\affiliation{ \textit Department of Physics, City University of Hong Kong, Hong Kong}

\date{\today}

\begin{abstract}

Dynamical quantum phase transitions (DQPTs) are criticalities in the time evolution of quantum systems and their existence has been theoretically predicted and experimentally observed. However, how the system behaves in the vicinity of DQPT and its connection to physical observables remains an open question. In this work, we introduce the concept of the Loschmidt amplitude spectrum (LAS), which extends the Loscmidt amplitude - the detector of the transition - by considering the overlap of the initial state to all the eigenstates of the prequench Hamiltonian. By analysing the LAS in the integrable transverse-field Ising model, we find that the system undergoes a population redistribution in the momentum space across DQPT. In the quasiparticle picture, all the lower-half k-modes are excited when the system is at DQPT. The LAS is also applicable to study the dynamics of non-integrable models where we have investigated the Ising model with next-nearest-neighbour interactions as an example. The time evolution of the system’s magnetization is found to be connected to the products of the LAS and there exists a simultaneous overlap of the time-evolved state to pairs of eigenstates of the prequnech Hamiltonian that possess spin configurations of negative magnetization. Our findings provide a better understanding of the characteristics of the out-of-equilibrium system around DQPT.

\end{abstract}

\maketitle

\section{Introduction}

The research on dynamical quantum phase transitions (DQPTs) has been thriving, both in experimental \cite{Oodptiatns,DOoDQPTiaIMBS,Ooambdptwa53qqs,Oodvaqiaswt,OoaDQPTbaSCS} and theoretical researches \cite{DQPTitTFIM,DQPTiSCwLRI-MDCoNC,PGSPTtQD,Dpdoqscwlri,DQPTaQPE,DqptitKhm,Qditfimfcn,Dotesisc,Doeeaescaqpt,SaUaDQPT,Bbcfdptiodtias,Neqsdfcsp,DEaDQPTiICvOoTOC,DQPTiSwBSP,Dtopffe,Cadqptiaiti,DTTitMSMwatT,Ddptfept,Dqpticarooaqs,DqptiWs,Fdqpt,Dqptidtc,DqptanMd,Ndoaswtkofaap,Qdazem:TcotCm,Fdptaes,FdqptiteXYm:Ntatt,ODQPTtQSO,Lmodqpt,Lmfiti,Adpiqspwlri,Dqptfacqq,Dptaqinm,Dqptitannnic,Cdqpt}, among condensed matter physicists . On the one hand, quantum simulators have allowed access to the real-time dynamics of quantum systems in experiments \cite{Oodptiatns,DOoDQPTiaIMBS,MEGiQDoBiaOL,Necdiodbg,Nlpociqswlri,Ooambdptwa53qqs,Oodtcoiaddmbs,Oodvaqiaswt,Pmbdoa51aqs,OoaDQPTbaSCS,PtQDoACia2DQISS,PtrteiaiscodBg,QCiaIC-EEfEE8S,Qeaepiaqmbs,RaPiaIQS,UDQSwTI}, among which detection of DQPT was found possible \cite{Oodptiatns,DOoDQPTiaIMBS,Ooambdptwa53qqs,Oodvaqiaswt,OoaDQPTbaSCS}. On the other hand, theoretical studies of DQPTs have advanced our understanding of non-equilibrium physics in quantum many-body systems, among which may lead to potential applications in quantum computing \cite{MBLaTiQSM,Rpimbl}.

The central study of DQPTs relies on the concept of \textit{Loschmidt amplitude} (LA), which measures the overlap of the time-evolving system onto its initial state, i.e. 
\begin{equation}   \label{LA}
	\mathcal{G}_0(t)   =   \bra{\psi_0(g_i)} e^{-iH(g_f)t} \ket{\psi_0(g_i)},
\end{equation}
where $\ket{\psi_0(g_i)}$ is the ground state of the Hamiltonian $H(g_i)$ and $H(g_f)$ is the quenched Hamiltonian. DQPTs are defined by the zeros in the \textit{Loschmidt echo} (LE) $   \mathcal{L}(t)   =   |\mathcal{G}(t)|^2   $ or the non-analyticities in its rate function $\lambda(t)=-\lim_{N\to\infty}\ln\mathcal{L}(t)/N$, where $N$ is the system size  \cite{DQPTitTFIM}. They are analogous to the zeros in the partition function and the non-analyticities in the free energy in equilibrium phase transitions, and therefore $\lambda(t)$ is also called the dynamical free energy. Previous studies show that in most cases, DQPTs occur when the system is quenched across its equilibrium critical point $   g_c   $ \cite{Dqpt-ar,DqptRA} though there are exceptions found in, for examples, Ref. \cite{DQPTaQPE,DqptitKhm}.  The above are in fact type-II DQPTs. There are also type-I DQPTs describing the order parameter in late time staying finite or vanishing in different dynamical phases \cite{DQPTiSCwLRI-MDCoNC,RaPOotOPiFC,QQaOEDTitIDBHM,Dpdoqscwlri}. A recent studies has shown that the two types of DQPTs are actually related in the long-range transverse field Ising model \cite{DQPTiSCwLRI-MDCoNC}. In this paper, we will focus on type-II DQPTs.

In spite of the wealth of literature successfully arguing the presence of DQPT in different models \cite{Dqpt-ar,DqptRA}, the question of how the system behaves in the vicinity of DQPT is yet to be addressed. Various physical quantities have been investigated in attempt to unveil the system's characteristic around DQPT. For examples, nonanalytical behaviors are found in the correlation matrices and crossings and degeneracy in the entanglement spectrum are observed at DQPTs \cite{Qditfimfcn,Dotesisc,Doeeaescaqpt,SaUaDQPT,Bbcfdptiodtias,Fdptaes}. It is also showed that there exists correspondence between DQPT and the systems' equilibrium order parameter in some models. For instance, the magnetization in the transverse-field Ising model and its variations was found both analytically and experimentally switching between positive and negative regime at the critical times \cite{DQPTitTFIM,DOoDQPTiaIMBS,Neqsdfcsp,DEaDQPTiICvOoTOC,DQPTiSwBSP}, providing a more physical linkage of DQPTs to physical observables. Dynamical topological order parameters are also introduced to study the topological properties of DQPTs. The Pancharantnam geometric phase in the momentum space is found to exhibit a discontinuous jump at critical times \cite{Dtopffe,Oodvaqiaswt,Oodptiatns,Cadqptiaiti,Dqptfacqq} in non-interacting models. Another dynamical topological order parameter defined by the time-ordered two-point Green's function, which is applicable for interacting systems, is also found to have discontinuity across the DQPT \cite{DTTitMSMwatT}.

 In this work, we attempt to provide insights to the above-mentioned open question. Motivated by the observation that the dynamics of a quantum system shall depend on the whole spectrum of the Hamiltonian, we extend the conventional definition of the LA in Eq. (\ref{LA}), which just take into account the overlaps onto the ground state, to the overlaps onto the excited states. We name the extension the \textit{Loschmidt amplitude spectrum} (LAS) of a system. The LAS does not require any non-trivial knowledge of quantum dynamics such as the order parameters, rather one only needs to retrieve the spectrum of the system concerned, which is already computed upon computing the dynamical time-evolution of the initial state. The LAS is a conceptually lighter alternative to access the dynamic profile of a general condensed matter system.

 To show how one uses the LAS in practice, we applied the LAS to investigate the dynamics in both integrable and non-integrable models. We found that the integrable 1D transverse-field Ising model experiences temporary population migration in the momentum around dynamical phase transitions. Namely the quench triggers excitation concentrated on the middle range of the allowed \textit{k} values in the momentum space, whereas at DQPT the excitation shifts to the whole lower-half range.  We also examined the LAS  for various quench parameters, and confirmed that nonanalyticities persist in LAS for typical quenches such as quenches within one phase. On the other hand, the non-integrable 1D axial next-nearest-neighbor Ising model exhibits substantial drop of magnetization when quenched from paramagnetic (PM) phase to ferromagnetic (FM) phase, and we found that the drop can be attribute to the simultaneous finite overlap between the time-evolved state and eigenstate pairs with negative spin magnetisation of the prequenched Hamiltonian.

The paper is organized as the following: The definition of LAS is presented in Sec. \ref{sec:LAS}. The results for 1D transverse-field Ising model, including the analytical expression of LAS in momentum space and numerical calculations of different quenches are given in Sec. \ref{sec:TFIM}. In Sec. \ref{sec:ANNNI}, the LAS of the 1D axial next-nearest-neighbor Ising model and its relation to the magnetization of the system are investigated. Finally, a conclusion is given in Sec. \ref{sec:conclusion}.

\section{Loschmidt amplitude spectrum}
\label{sec:LAS}

Consider a system described by the Hamiltonian $   H(g)   $ such that $   H(g)\ket{\psi_n(g)}   =   E_n(g)\ket{\psi_n(g)}   $, where $   \ket{\psi_n(g)}   $ is the \textit{n}th eigenstate of the Hamiltonian with the corresponding eigenenergy $   E_n(g)   $. Unless otherwise specified, we prepare the initial state of the system to be the ground state of the Hamiltonian at $   H(g_i)   $, and quench the system with $   H(g_f)   $ at the time $   t   =   0   $. The LAS is defined by
\begin{equation}   \label{LAex}
	\mathcal{G}_n(t)   =   \bra{\psi_n(g_i)} e^{-i H(g_f) t} \ket{\psi_0(g_i)},
\end{equation}
and the respective rate function spectrum
\begin{equation}   \label{ratefuncex}
	\lambda_n(t)   =   \lim_{N \rightarrow \infty} -\frac{1}{N} \log[\mathcal{L}_n(t)]
\end{equation}
where $   \mathcal{L}_n(t) = |\mathcal{G}_n(t)|^2  $ is the Loschmidt echo spectrum. In the followings, we use Loschmidt amplitude/echo/rate spectrum (LAS/LES/LRS) interchangeably.

For a general \textit{n}, the LAS measures the overlap between the time-evolving state and the \textit{n}th eigenstate of the initial Hamiltonian. It quantifies how much the quenched state is scattered into an excited state of the initial Hamiltonian. 
The quantity in Eq.(\ref{LAex}) represents the first column of the evolution operator matrix $e^{-iH(g_f)t}$ expanded in the eigenstates of $H(g_i)$ while the conventional definition of LA in Eq. (\ref{LA}) only captures the first diagonal element of the matrix. Therefore, we expect more information about the dynamics of the system to be encoded in the LAS. By analyzing the LAS and the corresponding rate function spectrum, insights into the characteristics of the quantum state around the DQPT can be drawn. The similar quantity has also been studied in the context of fidelity spectrum in equilibrium QPTs \cite{FS-ATtPtPoaQP} and many-body localizations \cite{CfMBLDPT}, and here we extend the study to the time-dependent case. In the following, the LAS in one-dimensional transverse-field Ising model and axial next-nearest-neighbor Ising model will be analyzed.

\section{1D transverse-field Ising model}
\label{sec:TFIM}

The Hamiltonian of the transverse-field Ising model is given by
\begin{equation}   \label{TFIMH}
	H(g)   =   -J\sum_{j = 1}^{N} (\sigma_j^x \sigma_{j + 1}^x   +   g\sigma_j^z),
\end{equation}
where $   J   $ is the Ising coupling, $   g   $ represents the external magnetic field strength, $   \sigma_j^\alpha   $ ($   \alpha   =   x,y,z)   $ are the Pauli matrices of site $   j   $. The periodic boundary condition is adopted. Below we set $   J   =   1   $ for convenience.

The model can be diagonalized by the Jordan-Wigner transformation $   \sigma_j^+   =   \exp\big[ i\pi \sum_{n   =  1}^{j   -   1} c_n^\dagger c_n \big]c_j   $, to a spinless fermionic model, followed by a Fourier transformation $   c_j   =   (1/\sqrt{N})\sum_k e^{ikj} c_k   $, where the Hamiltonian becomes
\begin{equation} \label{TFIMHk}
	\begin{aligned}
		H(g)   =   & \sum_{k > 0} \big[ ( \cos(k)   -   g ) ( c_k^\dagger c_k   -   c_{-k} c_{-k}^\dagger ) \\
		& +   i \sin(k) ( c_k^\dagger c_{-k}^\dagger   -   c_{-k} c_k ) \big],
	\end{aligned}
\end{equation}
with $   c_k(c_k^\dagger)   $ being a set of fermionic annihilation (creation) operators with $   k   =   \pi/N,3\pi/N,\dots,(N   -   1)\pi/N  $ for even $N$ and $k=0,2\pi/N,\dots,(N   -   2)\pi/N$  for odd $N$ in the even parity subspace \cite{QPS2nd} . 

The quadratic Hamiltonian in Eq. (\ref{TFIMHk}) can be diagonalized by performing Bogoliubov transformation, namely $   c_k   =   u_k(g) \beta_k   +   i v_k(g) \beta_{-k}^\dagger   $, where $   u_k(g)   =   \cos(\theta_k(g))   $ and $   v_k(g)   =   \sin(\theta_k(g))   $. The $   \theta_k(g)   \in   [ 0 , \pi/2 ]   $ is called the Bogoliubov angle satisfying the condition $   \tan(2 \theta_k)   =   \sin(k)/(\cos(k)   -   g)   $. The resulting Hamiltonian is
\begin{equation} \label{diagTFIMH}
	H(g)   =   \sum_{k > 0} \varepsilon_k(g) ( \beta_k^\dagger \beta_k   -   \beta_{-k} \beta_{-k}^\dagger ),
\end{equation}
with $   \varepsilon_k(g)   =   2\sqrt{( \cos(k)   -   g )^2   +   \sin^2(k)}   $. The ground state is the vacuum state $   \ket{0(g)}   $ such that $   \beta_k \ket{0(g)}   =   0   $ for all \textit{k}. The excited states can be generated by creating pairs of opposite-momentum quasiparticles in different $   k   $ modes on the vacuum state. A quantum phase transition takes place when the magnetic field changes across the critical point $   g_c   =   1   $. The system transforms from a FM phase to a PM phase when \textit{g} increases from below $   g_c   $ and vice versa.

Taking the initial state as the ground state of $   H(g_i)   $, the LA of the system for a sudden quench $   g_i   \rightarrow   g_f   $ has an analytical expression derived by Silva \cite{SotWDoaQCSbQaCP}: Let $   \eta_k   $ and $   \gk   $ be the eigenmodes of Hamiltonian $   H(g_i)   $ and $   H(g_f)   $ respectively, one can easily write the transformation in between as $   \eta_k   =   U_k \gk   -   i V_k \gnkd   $ with
\begin{equation} \label{UkVk}
	\begin{aligned}
		U_k   & =   u_k(g_i) u_k(g_f) + v_k(g_i) v_k(g_f) \\
		V_k   & =   u_k(g_i) v_k(g_f) - v_k(g_i) u_k(g_f).
	\end{aligned}
\end{equation}
Thus the LA is given by
\begin{equation} \label{TFIMLA0}
	\begin{aligned}
		\mathcal{G}_0(t)   =   & \frac{e^{-i E_0(g_f) t}}{\mathcal{N}^2} \bra{0(g_f)} \bigg[ e^{-i \sum\limits_{k > 0} \frac{V_k}{U_k} \gnk \gk} \\
		& \times   e^{i \sum\limits_{k > 0} \frac{V_k}{U_k} e^{-i2 \varepsilon_k(g_f) t} \gkd \gnkd} \bigg] \ket{0(g_f)},
	\end{aligned}
\end{equation}
where $   E_0(g_f)   $ is the ground-state energy of $   H(g_f)   $ and $   \mathcal{N}   $ is the normalization factor.

The \textit{ground-state} rate function can be calculated as
\begin{equation}   \label{ratefunc0}
	\lambda_0(t)   \sim   -\frac{1}{N} \sum_{k > 0} \ln\bigg[ 1   +   T_k^4   +   2 T_k^2 \cos( 2 \varepsilon_k(g_f) t ) \bigg],
\end{equation}
where $   T_k   =   V_k/U_k   =   \tan(\phi_k)   $ with $   \phi_k   =   \theta_k(g_i)   -   \theta_k(g_f)   $, and \textit{N} specifies the system size. Note that we have ignored an irrelevant constant term.
 Using Eq.(\ref{LAex}), we obtain the LAS 
\begin{align}   \label{LAkp}
	\mathcal{G}_n(t)   =   & \frac{e^{-i E_n(g_f) t}}{\mathcal{N}^2} \prod_{\kp} \bigg( 2T_{\kp}\sin( \varepsilon_{\kp}(g_f) t ) e^{-i\varepsilon_{\kp}(g_f) t} \bigg) \nonumber \\
	& \times   \prod_{k \neq \kp > 0} \bigg( 1   +   T_k^2 e^{-i2\varepsilon_k(g_f)t} \bigg),
\end{align}
where the $   \kp   $ product includes all occupied $   k   $ states, $   E_n(g_f)   $ is the energy of the corresponding excited state. With the expression above, we can compute the Loschmidt rate easily
\begin{equation}   \label{ratefunckp}
	\begin{aligned}
		\lambda_n(t)   & \sim   -\frac{1}{N} \bigg\{ \sum_{\kp} \ln\bigg[ 2T_{\kp}^2 ( 1   -   \cos( 2\varepsilon_{\kp}(g_f) t ) ) \bigg] \\
		& +   \sum_{k \neq \kp > 0} \ln\bigg[ 1   +   T_k^4   +   2T_k^2 \cos( 2\varepsilon_k(g_f) t ) \bigg] \bigg\}.
	\end{aligned}
\end{equation}

Comparing Eq. (\ref{ratefunckp}) with Eq. (\ref{ratefunc0}), they are similar to each other but the former one contains an extra term $\sum_{k'}\Lambda_{k'}(t)$, where 
\begin{equation}   \label{logkp}
	\Lambda_{\kp}(t)   =   -\frac{1}{N} \ln\bigg[ 2T_{\kp}^2 ( 1   -   \cos( 2\varepsilon_{\kp}(g_f) t ) ) \bigg].
\end{equation}

Further to the LAS, one can generalize to derive an analytical expression of the overlap between the time-evolving \textit{n}th eigenstate and the \textit{m}th eigenstate, namely
\begin{equation}   \label{Gmn}
	\mathcal{G}_{mn}(t)   =   \bra{\psi_m(g_i)} e^{-iH(g_f)t} \ket{\psi_n(g_i)}.
\end{equation}
If resolved in Bogoliubov eigenbasis one obtains two distinct forms. The first is the diagonal term where $m=n$,
\begin{equation}
	\begin{aligned}
		\mathcal{G}_{nn}(t)   =   & \frac{e^{-iE_0(g_f)t}}{\mathcal{N}^2} \prod_{k'} \bigg( T_{k'}^2   +   e^{-i2\varepsilon_{k'}(g_f)t} \bigg) \\
	& \times   \prod_{k   \neq   k'   >   0} \bigg( 1   +   T_k^2e^{-i2\varepsilon_k(g_f)t} \bigg),
	\end{aligned}
\end{equation}
where it gives the exact same LE and in turn LR as that for the ground state. The other form corresponds to the case when $   m   \neq   n   $ and is given by
\begin{equation}
	\begin{aligned}
		\mathcal{G}_{mn}(t)   =   & \mathcal{A}(t) \prod_{k'} \bigg[ iV_{k'}\big( U_{k'}   +   V_{k'}T_{k'} \big) (1   -    e^{-i2\varepsilon_{k'}(g_f)t} ) \bigg] \\
	& \times   \prod_{k   \neq   k'   >   0} \bigg( 1   +   T_k^2e^{-i2\varepsilon_k(g_f)t} \bigg),
	\end{aligned}
\end{equation}
where $   \mathcal{A}(t)   =   \frac{e^{-iE_0(g_f)t}}{\mathcal{N}^2}   $ is the insignificant prefactor. One can show that the corresponding Loschmidt echo reduces to the previously solved LES and thus the same rate function as Eq. (\ref{ratefunckp}). In other words, we expect the LAS will be the same if we take the excited state as the initial state. In this paper, we would focus on the analysis of $   n   =   0   $ case where the initial state is taken as the ground state.

The term $\Lambda_{\kp}$ becomes non-analytic when the argument of the logarithmic function is zero. The associated ``critical time" is given by
\begin{equation}   \label{critTkp}
	t_m(\kp,g_f)   =   \frac{m \pi}{\varepsilon_{\kp}(g_f)}   \qquad   m =   1,2,3,\dots,
\end{equation}
which depends on the value of $   \kp   $ and $   g_f   $. For a general excited state, since $k'$ are independent, the critical times in the LRS will be determined by all the $k'$ modes in the excited state and their associated critical times given by Eq. (\ref{critTkp}). In fact, in the complex time plane, one easily realizes that the zeros of $   \mathcal{G}_n(z)   $, where $z\in \mathbb{C}$, are all lying on the imaginary time axis with a magnitude equal to $   t_m(k',g_f)   $. This implies that the nonanalyticities of LRs for excited states are insensitive to system size and they persist in large systems, unlike the case for ground-state rate function where it was shown in Ref. \cite{DQPTitTFIM} that the Fisher zeros cross the imaginary time axis at thermodynamic limit. However, one shall be reminded that the LAS defined in Eq. (\ref{LAex}) is not a simple analogy to partition function in statistical mechanics.

\begin{figure} [t!]
	\centering
	\includegraphics[width=8cm]{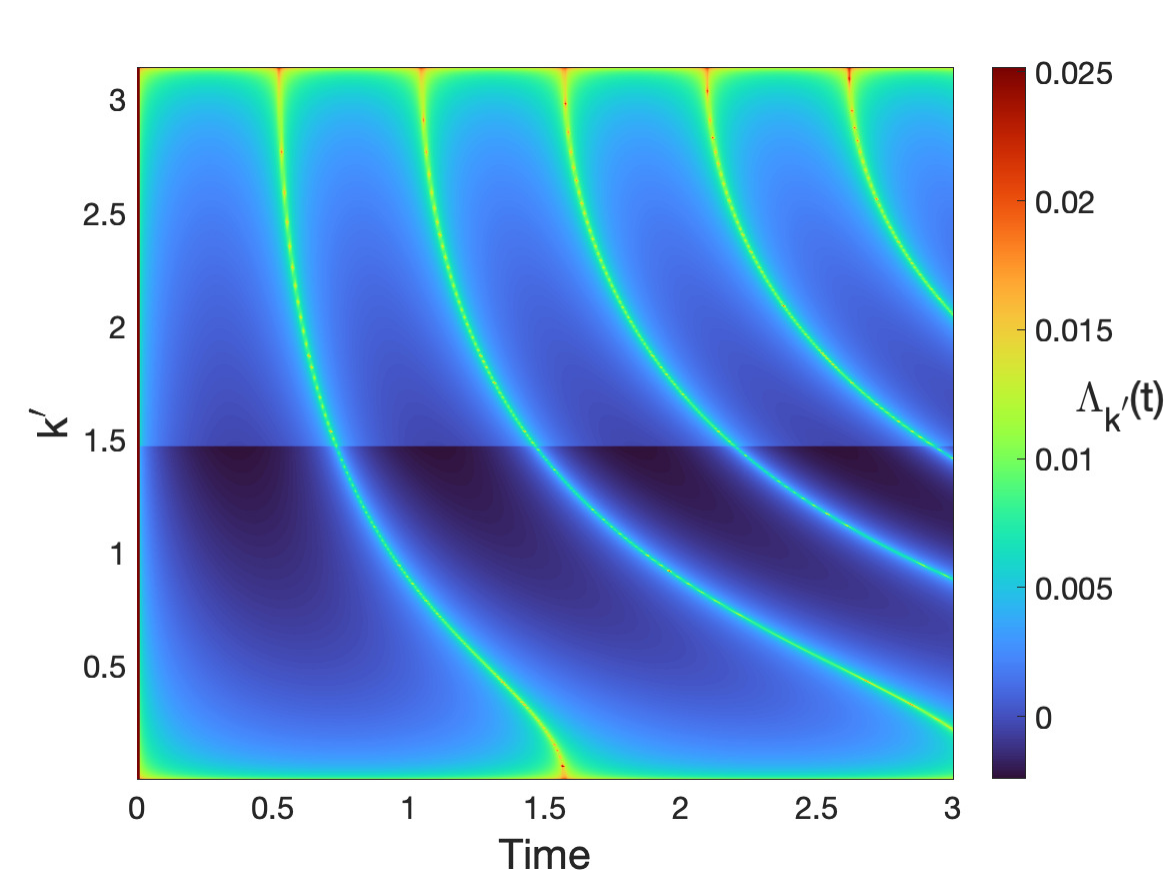}
	\caption{Color map of $   \Lambda_{\kp}(t)   $ in the rate function from an $   N   =   1000   $ system quenched from $   g   =   0.1   $ to 2. The non-analytic peaks in  $   \Lambda_{\kp}(t)   $  become more evenly separated and denser in time as $k'$ increases.}
	\label{critTkpcolor}
\end{figure}

Figure \ref{critTkpcolor} shows the plot of the critical time in Eq. (\ref{critTkp}) versus  $   \kp   $, which also corresponds to the case of single mode excitation. As expected, the critical time $t_m$ decreases as $k'$ increases. Note that an obvious horizontal ``line" crosses around the middle of the graph. This refers to the discontinuity of $   \lambda_n(t)   $ against $   \kp   $, where the Bogoliubov angle difference $   \phi_k   $ jumps from strictly positive to strictly negative as $   k'   $ increases, making the log function drop abruptly in magnitude and causing the large gap.

The $   \Lambda_{\kp}(t)   $ alone has a neat property when we concern large systems. A simple analysis shows the vanishing of the term, namely $   | \Lambda_{\kp}(t) |   \lessapprox   \ln( N )   /   N   $ for $   \kp   $ close to $   0,\pi   $ and thus $| \Lambda_{\kp}(t) |   \rightarrow   0$ in thermodynamic limit. It also holds true for a general allowed value of $k'$ as suggested from finite size analysis of our numerical results. Consequentially, the rate function for excited states with only a few occupied momentum states would behave similarly to the ground state quantity during quenching, i.e. a large main peak equivalent to the ground-state nonanalyticity, except some small spikes can be seen along at times in (\ref{critTkp}) given by $   \Lambda_{\kp}(t)   $.


\begin{figure} [t!]
	\centering
	\includegraphics[width=9cm]{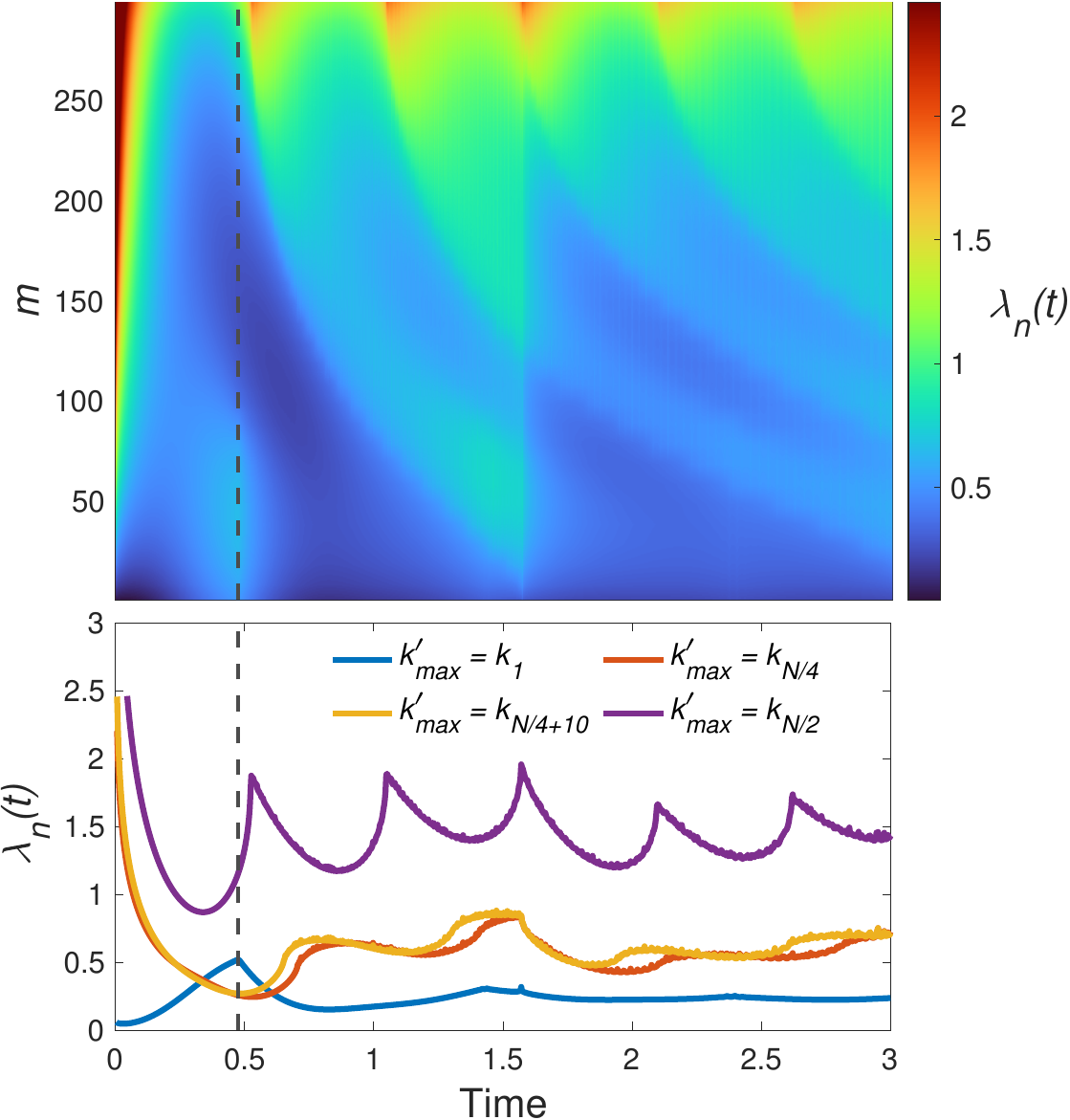}
	\caption{(Top) Variation of LRS for excited states of the form $   \prod_{\kp = k_1}^{\kp_{\textrm{max}}} \eta_{\kp}^\dagger \eta_{-\kp}^\dagger \ket{0(g_i)}   $, where $k_m=(2m   -   1)\pi/N$ with $m=1,2,\cdots,N/2$. (Bottom) Same quantities as the top panel with 4 specific values of $k'_{\max}$. The $   N   =   300   $ system is quenched from $   g   =   $ 0.1 to 2. Dashed black lines indicates the first critical time in the ground state rate function.}
	\label{nmodeex}
\end{figure}

On the other hand, the growth of the rate functions for higher excited states starts to behave differently and becomes more dramatic. The vanishing of $   \Lambda_{\kp}(t)   $ do not apply to states with multi-mode excitation \cite{SuppMat}. Figure \ref{nmodeex} illustrates the trend of the rate functions for $ \psi_n(g_i)= \prod_{\kp = \pi/N}^{\kp_{\textrm{max}}} \eta_{\kp}^\dagger \eta_{-\kp}^\dagger \ket{0(g_i)}   $, where the product is taken over all occupied $k$ modes, for a quench across the equilibrium critical point. Comparing the rate functions at the ground-state critical time, we can see the turning from nonanalytical peaks to smooth valleys, and then rise again to sharper peaks as we go up along the black dashed line.  The valleys in rate functions correspond to the high probability of the overlap of time-evolving state to the respective excited states. This suggests that the system is driven to a combination of states with lower-half momentum states being occupied at the DQPT when the system is quenched across the equilibrium critical point.

\begin{figure} [t!]
	\centering
	\includegraphics[width=9cm]{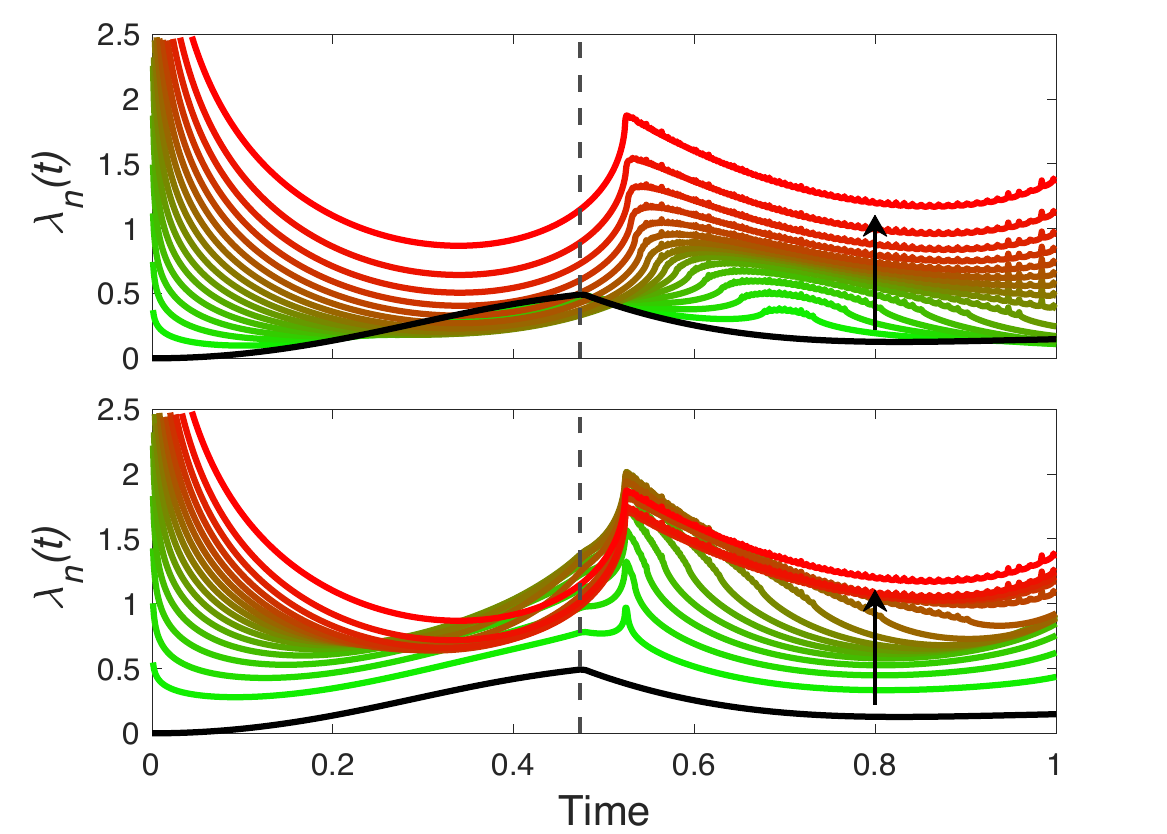}
	\caption{The rate functions for the excited states of the form $   \prod_{k'   =   k_{N/4   -   m}}^{k_{N/4   +   m   +   1}} \eta_{k'}^\dagger \eta_{-k'}^\dagger \ket{0(g_i)}   $ with $   m   =   0,1,2,\cdots,N/4   -   1   $ (top) and $   \prod_{k'   =   k_{N/2   -   m'}}^{k_{N/2}} \eta_{k'}^\dagger \eta_{-k'}^\dagger \ket{0(g_i)}   $ with $   m'   =   0,1,2,\cdots,N/2   -   1   $ (bottom) of an $   N   =   300   $ system quenched from $   g   =   0.1   $ to 2.0. Black curve represents the ground-state Loschmidt rate with dashed line indicating the first critical time, and black arrow shows the direction of increasing \textit{m}.}
	\label{nrightlastmodeex}
\end{figure}

\begin{figure} [t!]
    \centering
    \includegraphics[width=8cm]{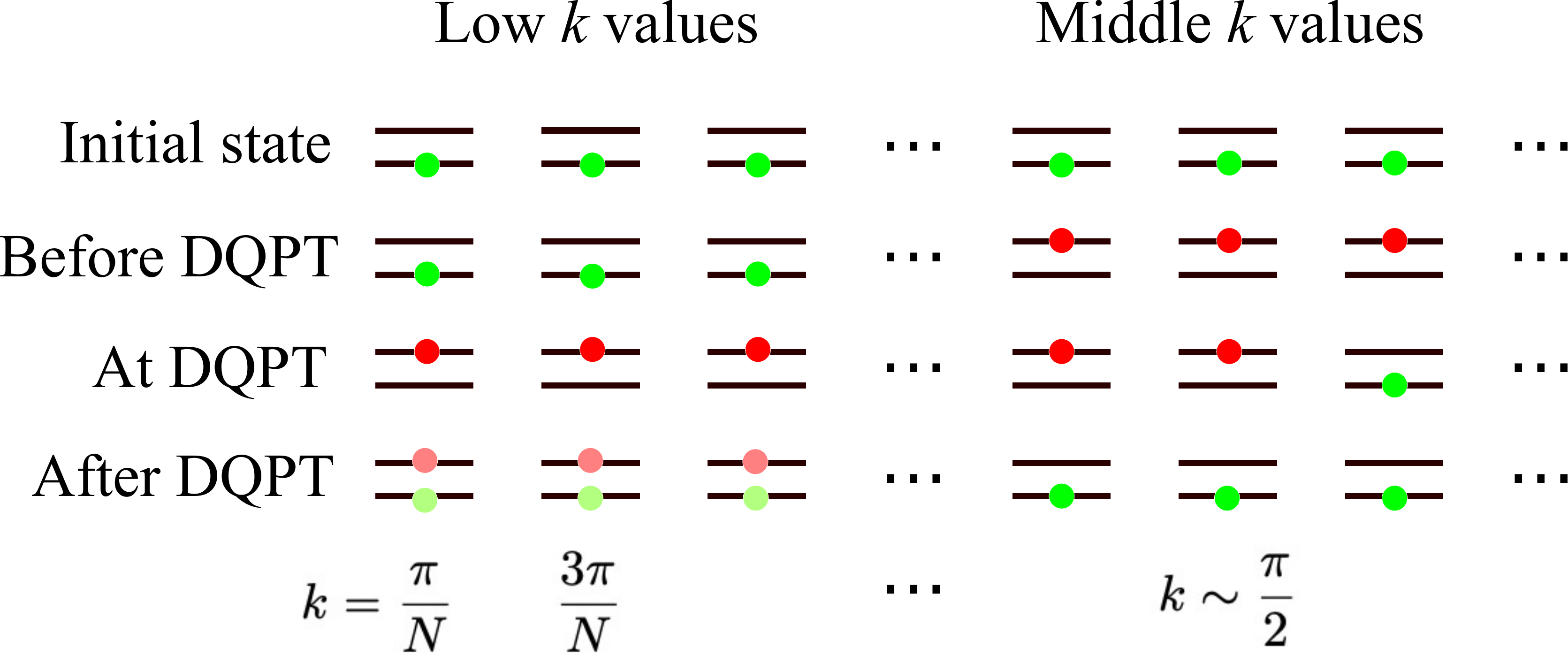}
    \caption{Schematic diagram of the DQPTs of 1D TFIM in the diagonalized space. The dynamics of quench around DQPT can be represented as a series of excitations among the two-level systems. Green and red dots correspond to the relaxed and excited quasiparticles respectively. The faded color dots represent a superposition of relaxed and excited quasiparticles at low \textit{k} values beyond critical time. Note that the quantum state of the system around the critical time is not only the state shown in the image, but a superposition of similar states. Subsequently the system would relax starting from the high \textit{k} modes.}
    \label{levelsdiag}
\end{figure}

A further diagnosis of the spectrum is presented in Fig. \ref{nrightlastmodeex}, where we explore higher excited states with excitations in the middle range of \textit{k} spectrum and excitations starting from the highest \textit{k} respectively. Overall the quenched system would barely stay in highly excited states and the excitations occur mainly in the middle or low \textit{k} modes during the dynamical evolution, as seen from Fig. \ref{nmodeex} and Fig. \ref{nrightlastmodeex}.

To summarize, Fig. \ref{levelsdiag} shows a schematic illustration of the dynamics of the system around DQPT. Notice that the following description is valid around the first critical time, and the long-term dynamics is out of the scope of our studies in this paper. The key concept is to realize that the diagonalized Hamiltonian in Eq. (\ref{diagTFIMH}) represents a system of \textit{N} two-level harmonic oscillators with independent momentum. The initial vacuum state refers to all the quasiparticles occupying the lower level. Once the system is being quenched, quasiparticles with momentum around the middle range of \textit{k} spectrum (i.e. $k\sim \pi/2$) are excited to the upper level. At DQPT, they drop to lower levels and are immediately followed by excitation of \textit{k} modes at lower-half \textit{k} values. At later times, all the excited modes will gradually relax and return back to lower level. The sudden occupation distribution shift is when dynamical quantum phase transitions occur. This phenomenon also occurs in DQPTs where the quench does not cross any underlying transition points in the XY model \cite{Ddptfept}. A brief discussion of the LAS of the XY model is given in the Supplemental Materials.

\begin{figure*} [t!]
	\centering
	\includegraphics[width=8cm]{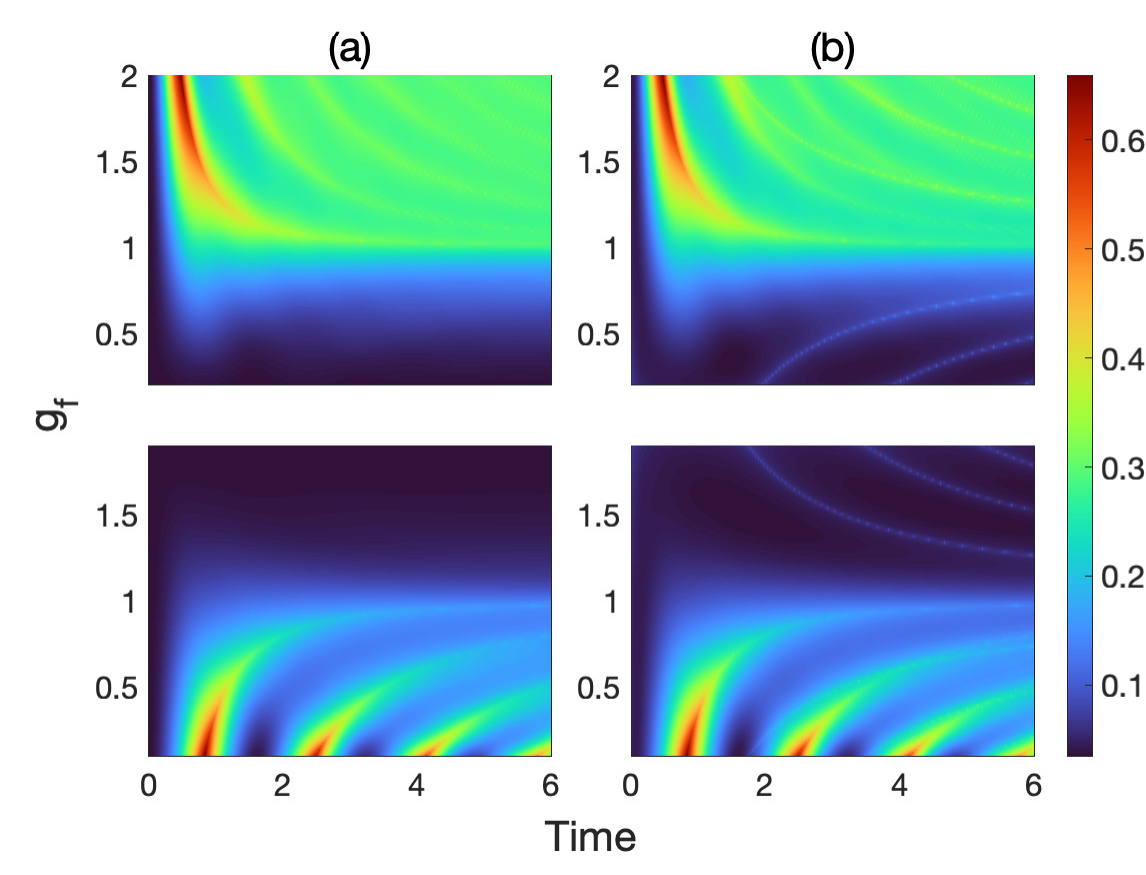}
	\includegraphics[width=8cm]{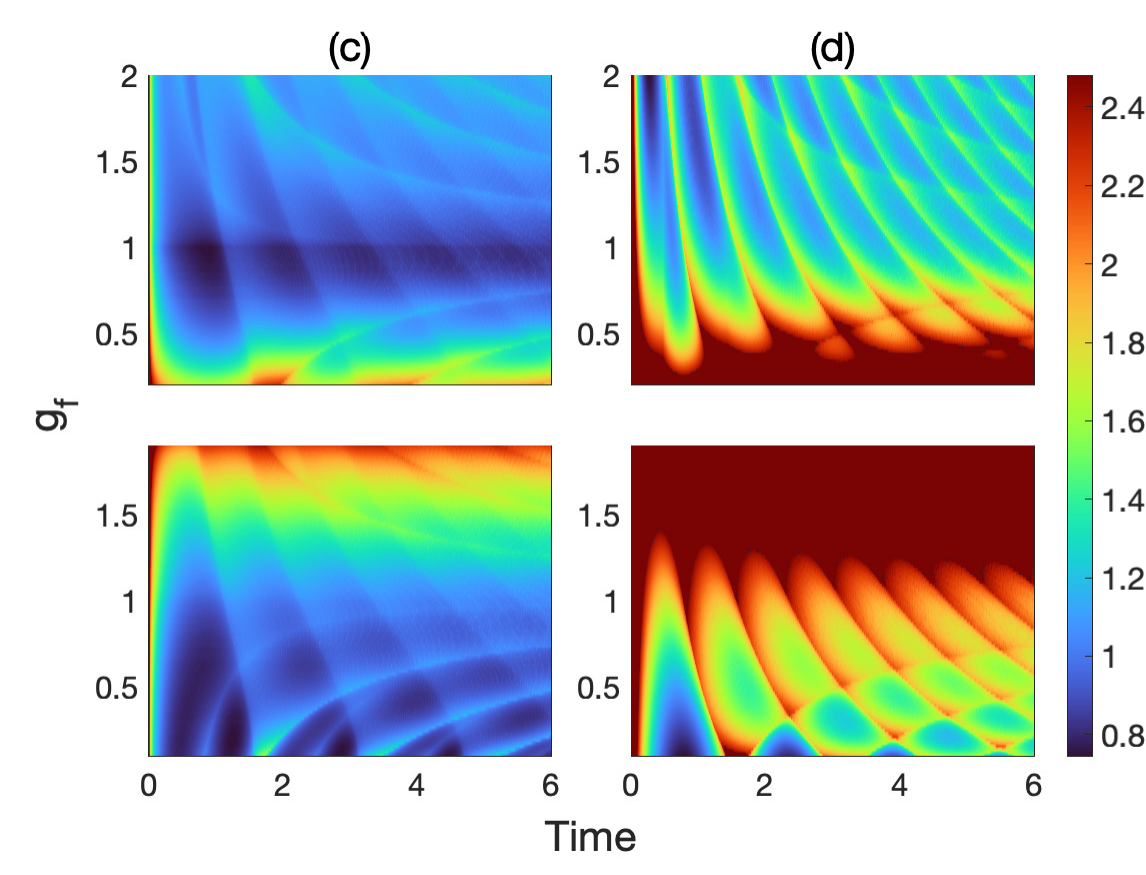}
	\caption{LRS with various final parameter, namely $   g_i   =   0.1   \rightarrow   g_f   $ from 0.2 to 2 for forward quench (top) and $   g_i   =   2   \rightarrow   g_f   $ from 1.9 to 0.1 for backward quench (bottom) for (a) around-state rate functions, (b) single-mode excited state, (c) lower-half excited state and (d) fully occupied excited state. Colors represent the intensity of $   \lambda_n(t)   $.}
	\label{tfimratefuncgfcolor}
\end{figure*}

The general quantum dynamics for TFIM is also studied. Figure \ref{tfimratefuncgfcolor} shows the rate functions for $\ket{\psi_n(g_i)}$ being the
\begin{enumerate}
    \item[(a)] ground state $   \ket{0(g_i)}   $;
    \item[(b)] 1-mode excited state $   \eta_{\kp}^\dagger \eta_{-\kp}^\dagger \ket{0(g_i)}   $;
    \item[(c)] lower-half excited state $   \prod_{\kp   =   k_1}^{k_{N/4}} \eta_{\kp}^\dagger \eta_{-\kp}^\dagger \ket{0(g_i)}   $;
    \item[(d)] fully excited state $   \prod_{\kp   =   k_1}^{k_{N/2}} \eta_{\kp}^\dagger \eta_{-\kp}^\dagger \ket{0(g_i)}   $,
\end{enumerate}
where $k_m=(2m   -   1)\pi/N$, for various quench parameters. There are obvious distinctions in the four cases. From figure \ref{tfimratefuncgfcolor} (a) and (b), we observed a divergence in the critical time (as indicated by the bright lines) at $   g   =   1   $, the equilibrium critical point separating the FM and PM phases. For the 1-mode excited state LR in figure \ref{tfimratefuncgfcolor}(b) , critical lines appear clearer to see where they are absent in the ground-state rate function. In addition, forward and backward quenches are fairly symmetric in case (a) and (b), whereas for higher excited states in case (c) and (d), this symmetry is broken, and the behavior of LRS becomes more dramatic that critical boundary starts to blur and kinks pass through the boundary. For the higher excited states, nonanalyticities are observed in the LRS for quenches within the same phase whereas the ground-state rate function goes smoothly as expected. The higher the excited state is, the denser the nonanalyticites in the LRS it possesses. It is still unclear that whether these nonanalyticities occurred in LAS follow the conditions to be a valid phase transition and would be an interesting topic for further studies.



\section{1D Axial next-nearest-neighbor Ising model}
\label{sec:ANNNI}

A variation of the quantum Ising model, also known as the axial next-nearest-neighbor Ising (ANNNI) model has also been shown to have exhibited DQPTs, and the next-nearest-neighbor interaction will further alter the characteristics of DQPT \cite{Dptaqinm,Dqptitannnic,Iowfoqdo1ds12annnim,Cdqpt}. In here we provide another point of view from the LAS to explain the dynamically critical phenomenon of the system. The Hamiltonian of ANNNI model is given by
\begin{equation}   \label{HANNNI}
	H(\Delta,g)   =   -J\sum_{j = 1}^{N} (\sigma_j^x \sigma_{j + 1}^x   +   \Delta \sigma_j^x \sigma_{j + 2}^x   +   g\sigma_j^z),
\end{equation}
with $   \Delta   $ controlling the next-nearest-neighbor interaction strength. When $\Delta=0$, the model is reduced to the quantum Ising model in Eq. (\ref{TFIMH}), where it can be diagonalized and quasiparticle picture interpretation applies. There are attempts to approximate transformation to retain quasiparticle picture using mean-field Jordan-Wigner transformation \cite{Iowfoqdo1ds12annnim}. Nonetheless, this work would present the numerical findings. The equilibrium ground state phase diagram of the model consists of four phases - the PM phase, FM phase, an antiphase (AP) phase with spin configuration of the form $|\uparrow\uparrow\downarrow\downarrow\uparrow\uparrow\cdots\rangle$, and an intermediate floating phase between the PM and AP phases \cite{Dptaqinm}. In the following, we consider quenching the system between the PM and the FM phase, where the phase boundary for $\Delta<0$ is given by $1+2\Delta=g_c+g_c^2\Delta/[2(1+\Delta)]$ \cite{Dptaqinm}. We show that the LAS can give insights into the magnetic property of the system.

\begin{figure*} [t!]
	\centering
	\begin{tabular}{cc}
		\includegraphics[width=9cm]{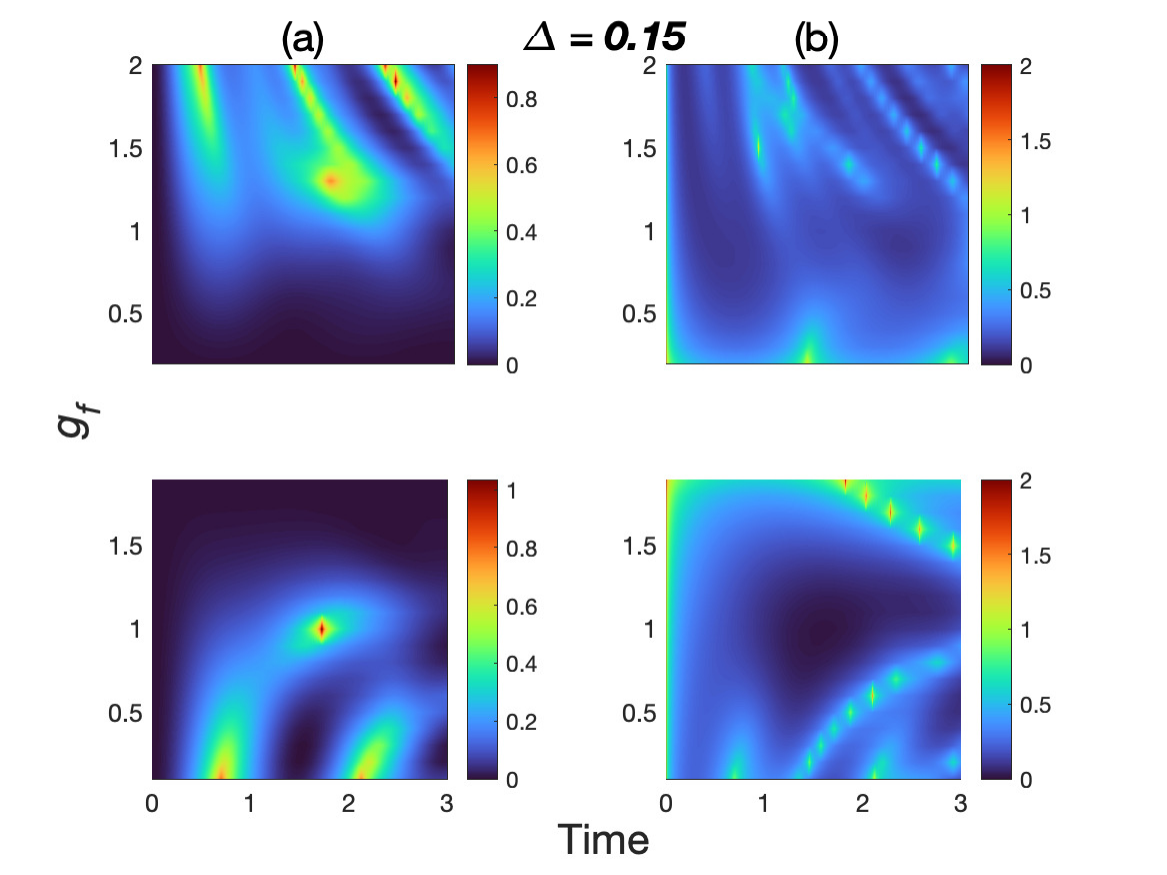}
		&
		\includegraphics[width=9cm]{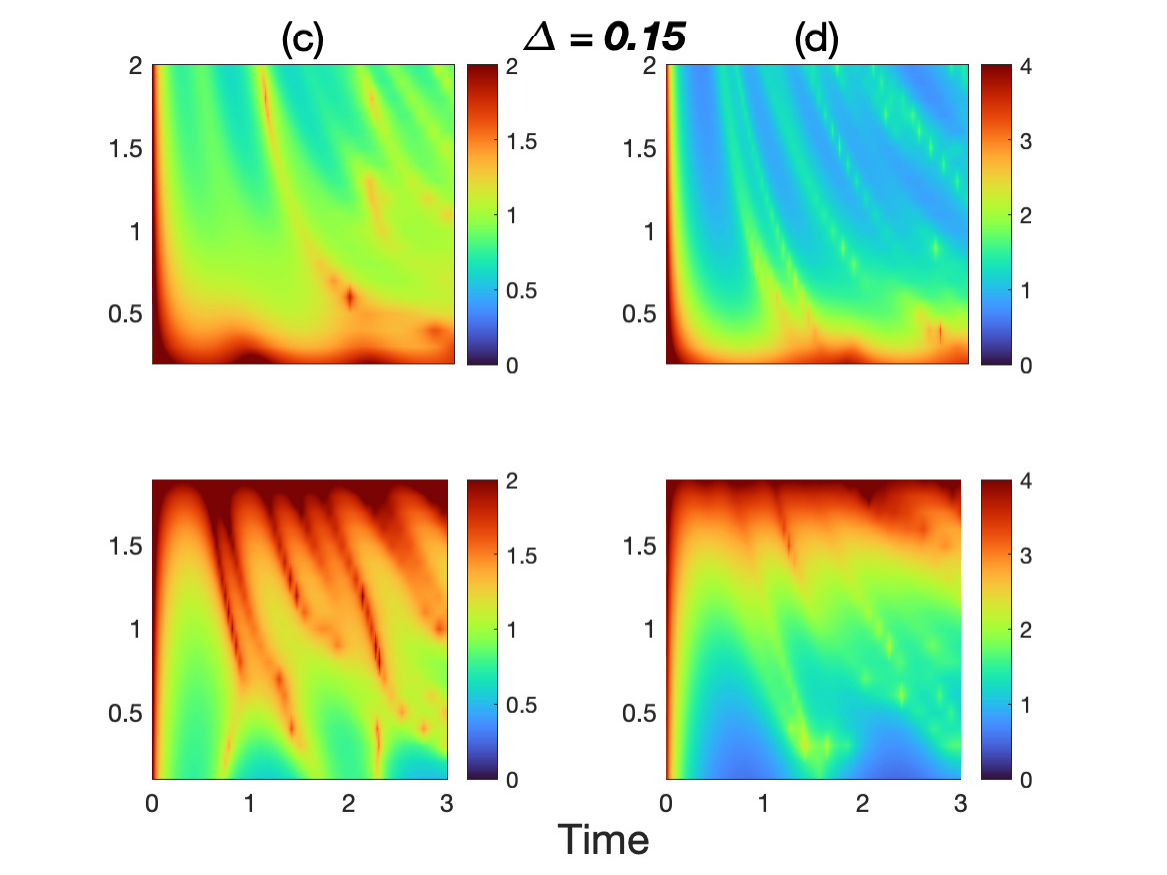}
		\\
		\includegraphics[width=9cm]{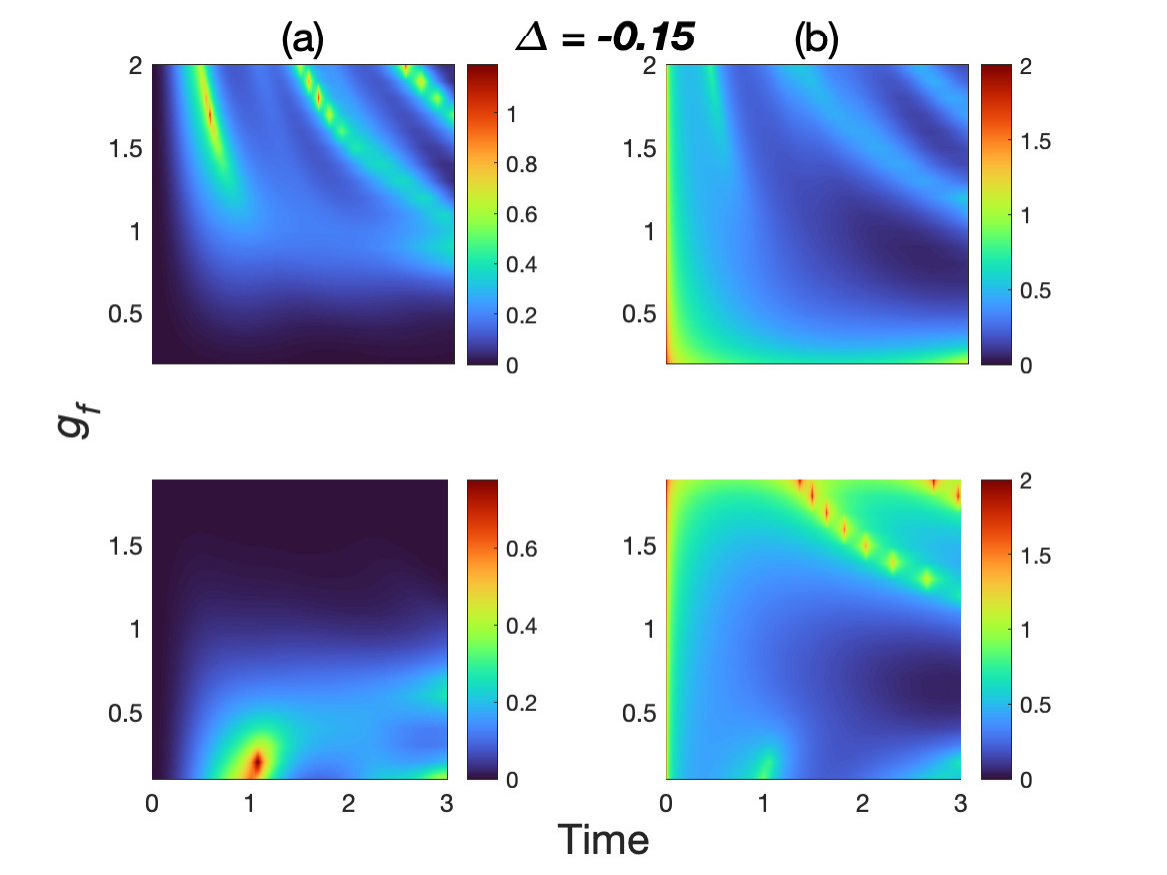}
		&
		\includegraphics[width=9cm]{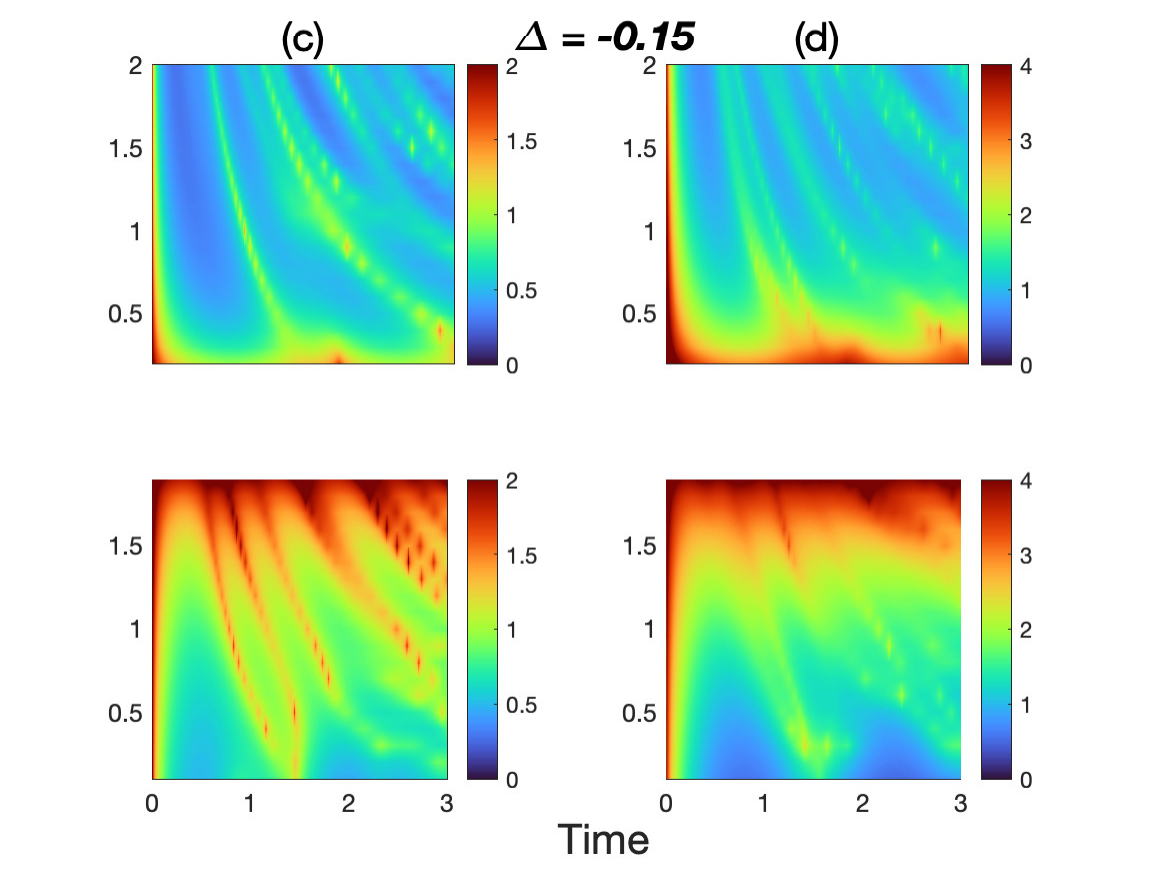}
	\end{tabular}
	\caption{Same colorplots as Fig. \ref{tfimratefuncgfcolor} for ANNNI model. The rows from the top to the bottom show the quench case for $(g,\Delta)=(0.1,0.15)\rightarrow(g_f,0.15)$, $(2.0,0.15)\rightarrow(g_f,0.15)$, $(0.1,-0.15)\rightarrow(g_f,-0.15)$, $(2.0,-0.15)\rightarrow(g_f,-0.15)$, respectively. The columns corresponds to the LRS for (a) the ground state, (b) the first excited state, (c) the middle-energy excited state and (d) the highest energy excited state, respectively.}
	\label{annniratefuncgfcolor}
\end{figure*}

We first analyze the general quench dynamics for ANNNI model as we did in the previous section. We focus on quench between the PM and FM phase in the following. The same colorplots as for TFIM are displayed in Fig. \ref{annniratefuncgfcolor} where the LRS for (a) ground state, (b) first excited state, (c) excited state of energy right in the middle of the spectrum and (d) highest excited state are considered for $\Delta=0.15$ and $\Delta=-0.15$ respectively. From the plots we see the similarities for the two models originating from the same universality class. Note that for small system size, the excited-state rate functions could behave very differently among each other compared to those for larger system size in the case of Ising model. 
Nevertheless, the general features can still be seen. 
The streamlined green peaks for the low energy states in Fig \ref{annniratefuncgfcolor}(a) and (b) approach the corresponding critical points $g_c(\Delta)$ asymptotically for both $   \Delta   $'s. As for the higher excited states, the nonanalytical peaks cross the underlying equilibrium phase boundary and the overall magnitudes are higher than that for lower energy states (Fig \ref{annniratefuncgfcolor}(c) and (d)). The effect of the NNN interaction can also be seen from plots. Namely, for the LR of the low energy states, the kinks for negative $   \Delta   $ are less prominent than that of positive $   \Delta   $, and the length between two consecutive kinks are slightly longer for negative $   \Delta   $ for the reason we will present below. On the other hand, the dynamics start to become ambiguous for higher energy states as seen in (c) and (d) in figure \ref{annniratefuncgfcolor}.

\begin{figure} [t!]
	\centering
	\includegraphics[width=8cm]{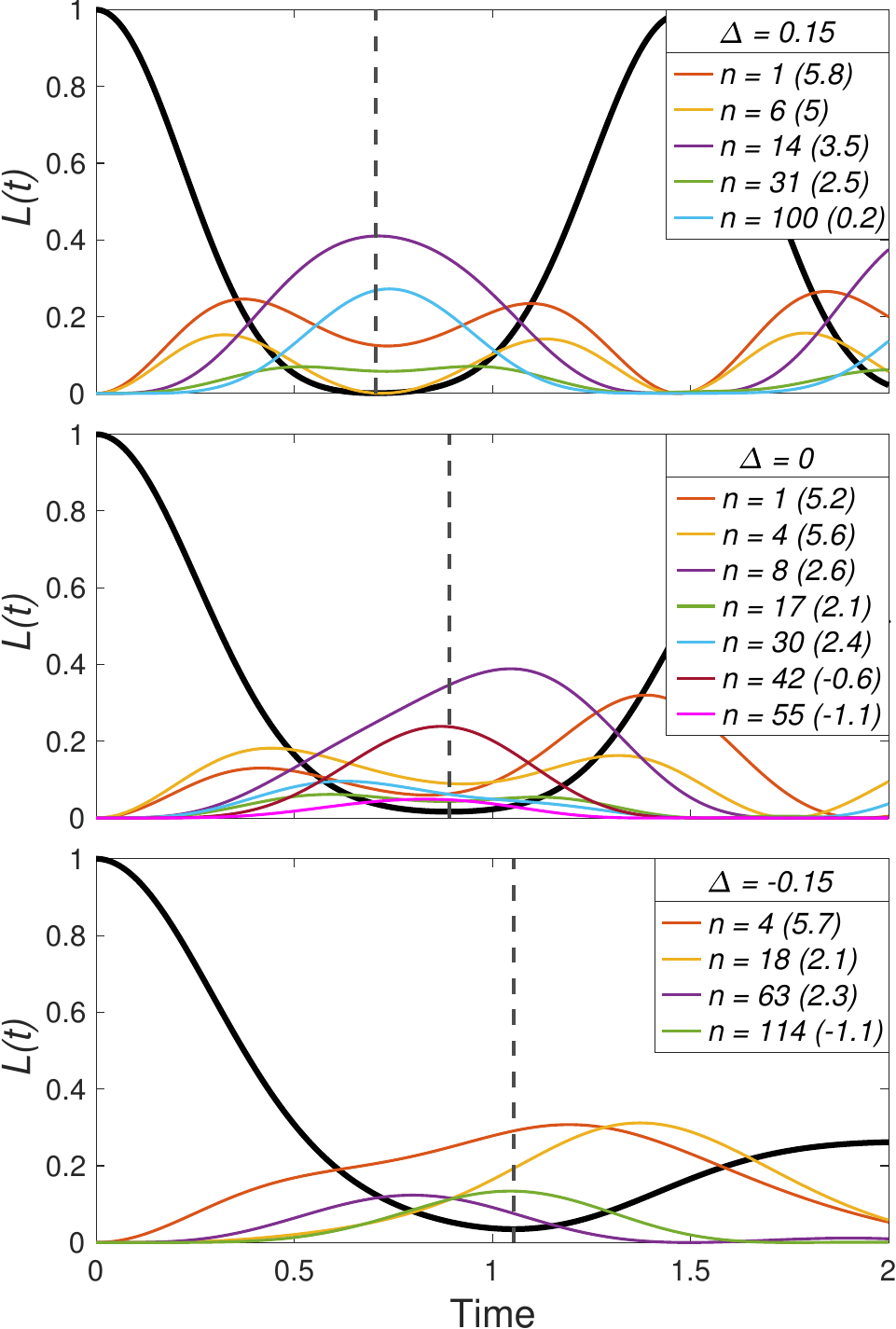}
	\caption{LES of the quenched $   N   =   10   $ ANNNI model from ordered phase to disordered phase with $   \Delta   =   0.15   $ (top), $   \Delta   =   0   $ (middle) and $   \Delta   =   -0.15   $ (bottom). The black curves correspond to the ground-state LE. The numbers in bracket refer to the magnetization of each involved excited state. Vertical dashed lines indicate the first critical time when DQPT happens.}
	\label{annniLEMzex}
\end{figure}

In the following we explore the physics during DQPT in ANNNI model by means of LAS. Figure \ref{annniLEMzex} shows LAS in a 10-site system with three different NNN interaction strength $\Delta$: -0.15, 0, 0.15. The $   \Delta   =   0   $ case refers to the TFIM and is plotted here for comparison. Only the eigenstates with relatively high contribution in the LAS and their corresponding magnetization are shown in the figure. Notice that all these highly contributed eigenstates are non-degenerate. From the figure, we observe a general DQPT process in the model as follows: For all considered values of $\Delta$, the system is first excited to low-lying excited states, followed by an increase of overlap to the higher excited states with lower magnetization around the vicinity of DQPT, then relaxes to lower energy states, restoring the high magnetization.

The effect of turning on the NNN interaction causes the system to lean on the eigenstates of weak magnetic character for positive $   \Delta   $, whereas it hinders the drop of magnetization and even prevents relaxation in the case of frustration where $   \Delta   $ is negative, aside from delaying DQPTs. This is the direct consequence of the NNN interaction being negative. The negativity of the interaction introduces a "frustration" on the spins, where the spins "hesitate" to align parallel to their nearest spins, but anti-parallel to their next-nearest neighbors to minimize energy. This "hesitation" multiplies during dynamical transition which collective spin flip takes place, so that spin flipping for negative $   \Delta   $ is reduced. The immediate effect would be the less number of effective states contributing to DQPT than the non-negative $   \Delta   $ cases and a later critical time as seen from Fig. \ref{annniLEMzex}.

\begin{figure} [t!]
	\centering
	\includegraphics[width=9.5cm]{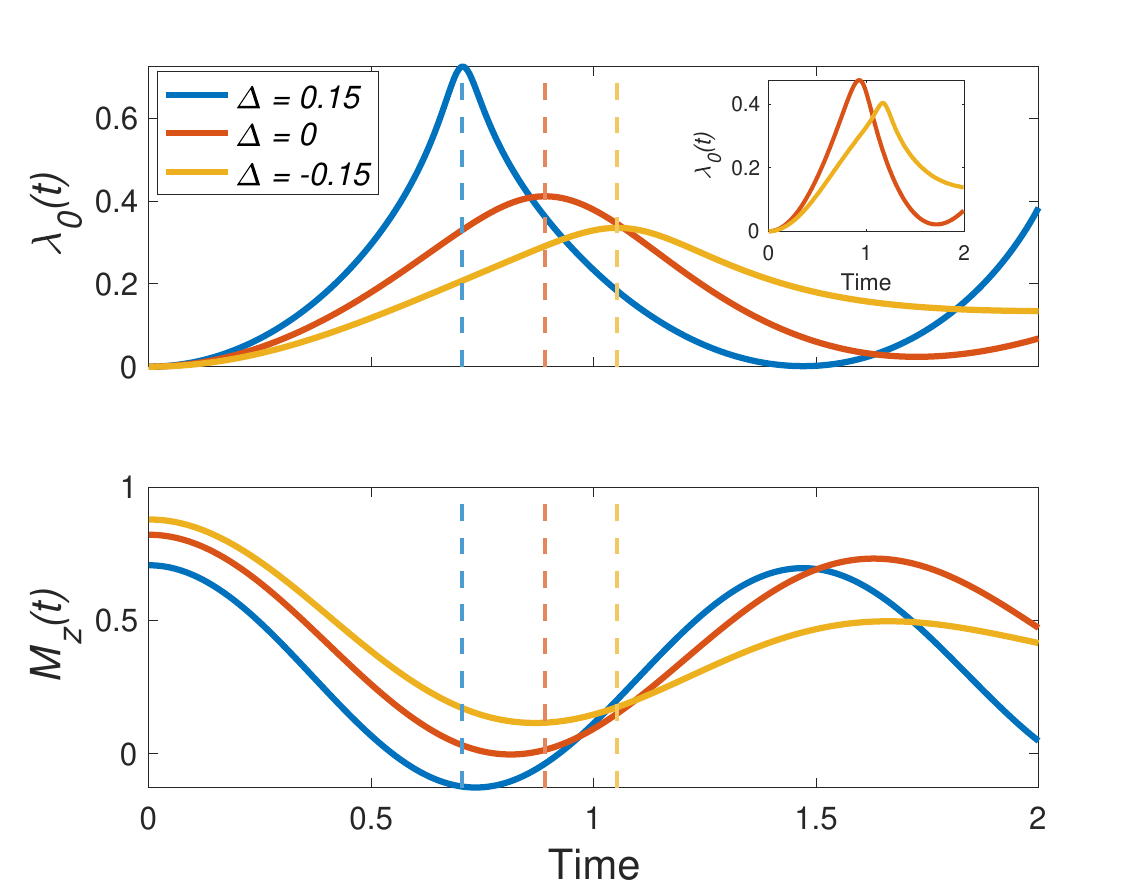}
	\caption{Quench dynamics and the corresponding time-varying magnetization of ANNNI model with 10 spin chain. The system is quenched from $   g_i   =   1.3   $ to $   g_f   =   0.2   $ with the corresponding critical times indicated by colored dashed lines. Inset shows the LRs for $\Delta=0$ and $-0.15$ for $   N   =   20   $.}
	\label{annnitimeMzex}
\end{figure}

LAS also reveals some surprising features in the magnetization dynamics. Figure \ref{annnitimeMzex} shows the ground-state rate function and the corresponding time-evolving magnetization as given by
\begin{equation}   \label{Mzt}
	\langle M_z(t) \rangle   =   \frac{1}{N} \sum_{j = 1}^N \bra{\Psi_i^f(t)} \sigma_j^z \ket{\Psi_i^f(t)},
\end{equation}
where $   \ket{\Psi_i^f(t)}   =   e^{-iH(\Delta,g_f)t}\ket{\psi_0(\Delta,g_i)}$.The peaks in $\lambda_0(t)$ become more prominent as system size increases (see inset of Fig. \ref{annnitimeMzex}), and be reminded that the peaks are true non-analytic for all $   \Delta   $'s shown in ref. \cite{Dptaqinm}. Note that each minimum of $   \langle M_z(t) \rangle   $ does not align strictly with the respective critical times, but qualitative importance can be well illustrated. As $\Delta$ increases from a negative value to a positive value, the time difference between the magnetization minimum and the rate function peak reduces. For a positive $\Delta$, the system's magnetization becomes negative in the vicinity of DQPT, while it stays positive for a negative $\Delta$.

\begin{figure*} [t!]
	\centering
	\hspace{-10pt}
	\includegraphics[width=6cm]{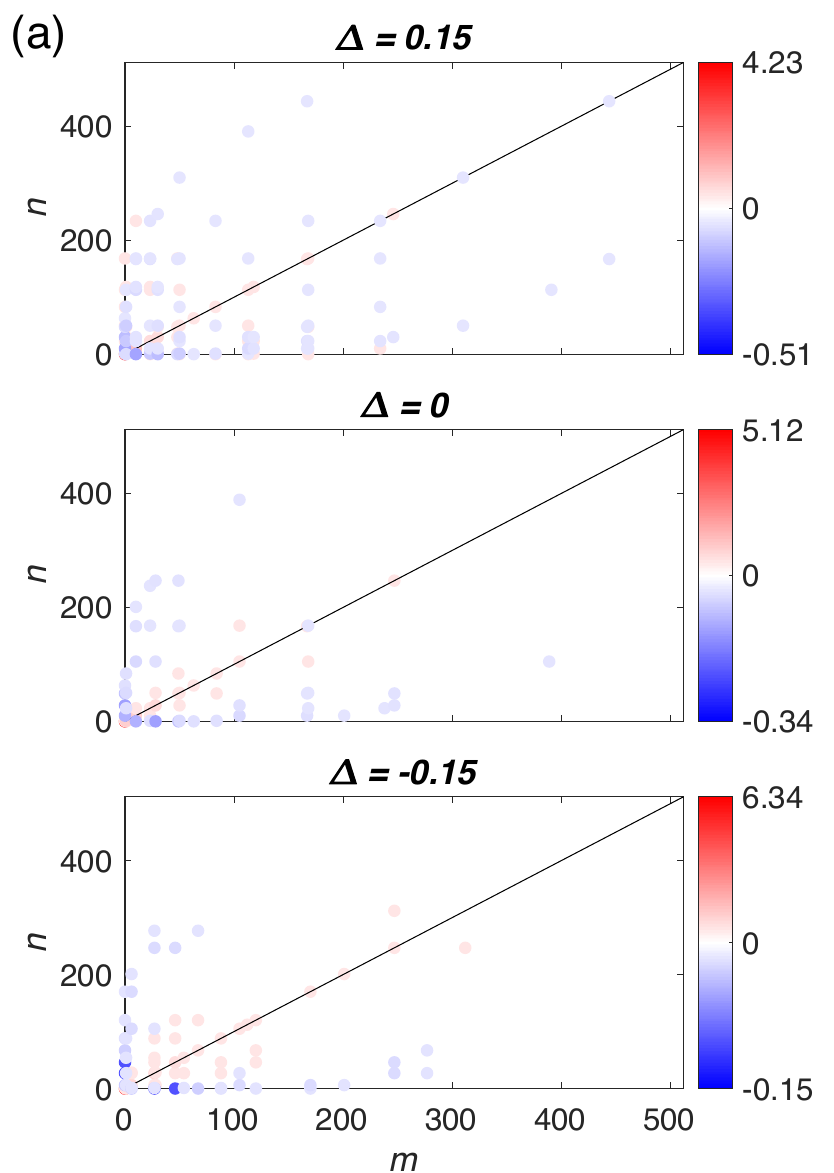}
	\hspace{-10pt}
	\includegraphics[width=6cm]{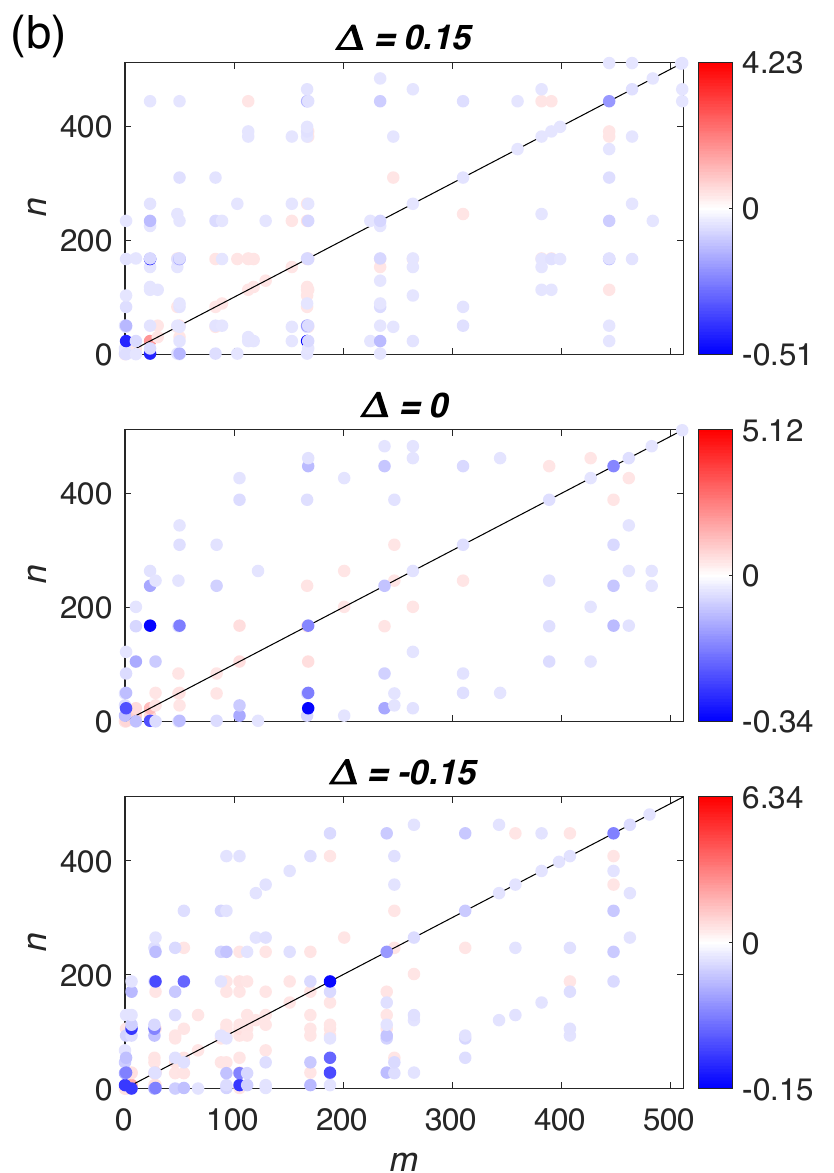}
	\hspace{-10pt}
	\includegraphics[width=6cm]{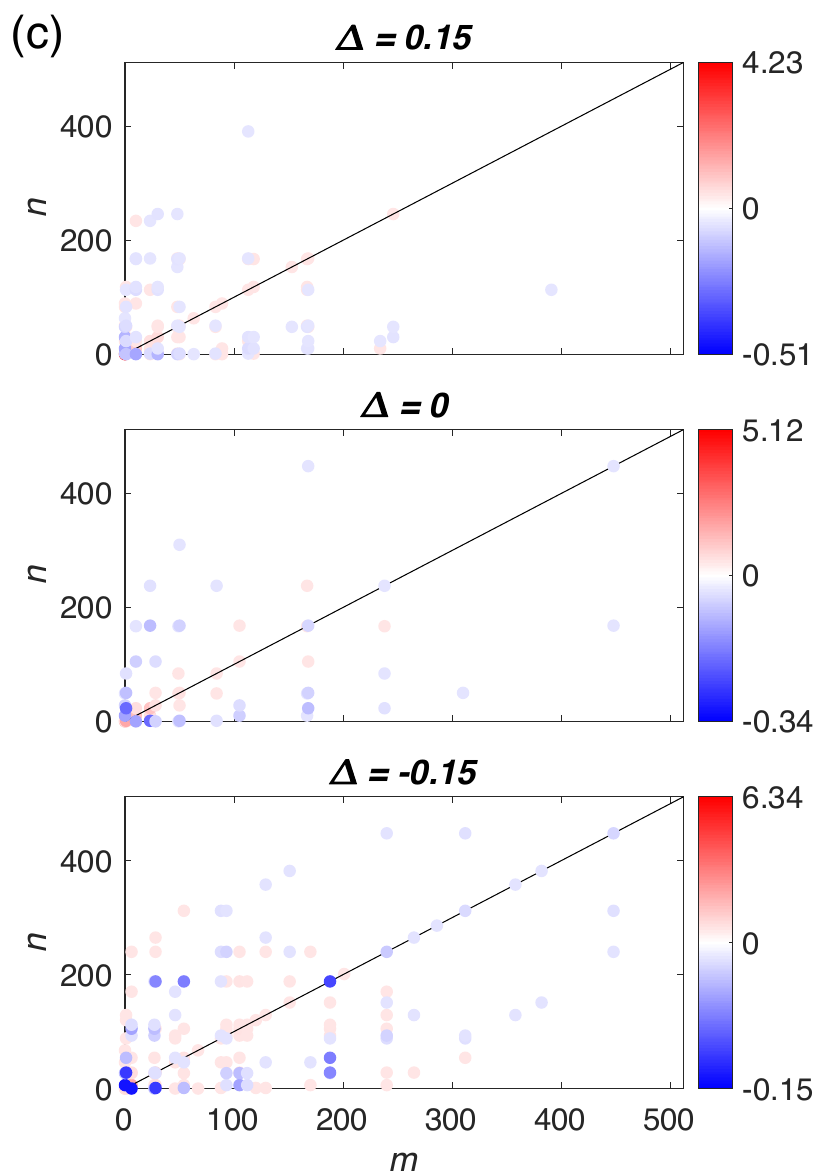}
	\hspace{-10pt}
	\caption{Explicit values of the term $   \mathcal{G}_m^*(t) \mathcal{G}_n(t) \bra{\Psi_m^i(0)} \sigma_j^z \ket{\Psi_n^i(0)}   $ for the three $   \Delta   $'s in three fixed times: (a) before DQPT, (b) at DQPT and (c) after DQPT. The quench is done on an $   N   =   10   $ system with quench pair $   (g_i,g_f)   =   (1.3,0.2)   $. Red color represents positive values and blue color represents negative values. Diagonal terms ($   m   =   n   $) are highlighted by the diagonal black lines.}
	\label{annniLAMzt}
\end{figure*}

To understand these, we can express $\langle M_z(t)\rangle$ in Eq. (\ref{Mzt}) in terms of the LAS as
\begin{equation}   \label{LAMzt}
    \avg{M_z(t)}   =   \frac{1}{N} \sum_{j = 1}^N \sum_{m,n} \mathcal{G}_m^*(t) \mathcal{G}_n(t) \bra{\Psi_m^i(0)} \sigma_j^z \ket{\Psi_n^i(0)},
\end{equation}
where $   \ket{\Psi_n^i(0)}   =   \ket{\psi_n(\Delta,g_i)}   $. In a sense, the magnetization is weighted by spectral behavior of the system. We plotted the individual terms in the summation of Eq. (\ref{LAMzt}) for different $   \Delta   $'s at different times in Fig. \ref{annniLAMzt}. Only real values are displayed since imaginary parts sum to zero. We observe the off-diagonal terms of $   M_z(t)   $ are finite and symmetric along diagonal as $   ( \mathcal{G}_m^*(t) \mathcal{G}_n(t) )^\dagger   =   \mathcal{G}_n^*(t) \mathcal{G}_m(t)   $, and they have great contribution to the time-evolving magnetization especially at critical time. Obviously $   \sigma_j^z   $ does not commute with the Hamiltonian, so $   \sigma_j^z   $ is not necessarily diagonal in initial eigenstate basis. However, since the matrix elements of $   \sigma_j^z   $ are fixed, it is the LAS $   \mathcal{G}_n(t)   $ guaranteeing some particular dynamical structure of the system during dynamical phase transition. Namely, the "active" off-diagonal terms are the most and most spread-out in the spectrum. At DQPT, the system is the most energetic that it would stay in various high-energy states, and preferrably superposition pairs of eigenstates $   [ \psi_m(g_i),\psi_n(g_i) ]   $ in a way to minimize the magnetization in \textit{z} direction. Note that most of the contributions come from the superposition pairs, and in the midst of the spectrum some pairs give the largest negative values. This is particularly true for positive $   \Delta   $, where the number of negative terms indicating as blue dots are more and possesses the lowest negative value among the three $   \Delta   $'s, as shown in the first plot of Fig. \ref{annniLAMzt}(b).

The spectrally weighted magnetization also shows distinguishable features away from critical time. Figure \ref{annniLAMzt}(a) and (c) captures the instants of $   \mathcal{G}_m^*(t) \mathcal{G}_n(t) \bra{\Psi_m^i(0)} \sigma_j^z \ket{\Psi_n^i(0)}   $ before and after the first DQPT respectively. In whichever time the effective terms concentrate on lower-energy spectrum (lower-left corner of the plots), while non-zero NNN interaction slightly triggers higher excitations. Imposing similar arguments as we described the mechanism of DQPT of TFIM, the quench stimulates the system through the lower spectrum, followed by the strongest superimposed state such that those pair states contribute, off-diagonally by calculation, to minimizing magnetization in our studied quench case. After DQPT, the system relaxed to low-lying states as presented by the vanishing of off-diagonal and higher-energy-state terms in Fig. \ref{annniLAMzt}(c). Note that the negative $\Delta$ case retains some more off-diagonal and higher-half spectrum terms because of the "frustration" explained above. 

\section{Conclusion}
\label{sec:conclusion}

In this work, we have introduced the LAS to investigate the physical nature of DQPTs in many-body systems. As examples we studied the LAS and the rate functions on 1D transverse-field Ising model and 1D axial next-nearest-neighbor Ising model. The former system displays a population re-distribution at the vicinity of dynamical phase transitions in the momentum space. In particular, the excitations originally concentrated in the middle range of $   k   $ values shift to the whole lower-half range at critical time, and relax to the lower levels after the transition.  We demonstrated too that LAS plays a role in the evolution of the observables in quenching, for instance the minimization of transverse magnetization in the study of ANNNI model. From that we infer that at DQPT the system, whether or not frustrated, tends to stay in excited states in pairs, in which their combined magnetic property achieve a vanishing or even flipped magnetization.

LAS is a conceptually simple but fundamental analysis method, where we directly diagnose the dynamics of the time-evolving systems without other complex physical treatments. This encourages its application to any other many-body systems. There will be interesting findings of how the emergence of LAS in DQPTs are presented in other quantum models, and even as a general sense of describing the physics underlying in DQPTs, where like the magnetization of ANNNI model studied in this work, any possible link of LAS to the system's order parameters can be established.

\begin{acknowledgments}
We thank Wen-Long You for helpful discussions. This work is financially supported by Research Grants Council of Hong Kong (Grant No. ECS/21304020) and City University of Hong Kong (Grant No. 9610438).
\end{acknowledgments}

\newpage

\appendix


\begin{widetext}
\section{Loschmidt amplitude spectrum in DQPT without crossing the equilibrium critical point}
\label{sec:xyLASDQPTnocross}

Dynamical phase transitions was also shown to occur in some models where the quench is without crossing the model's equilibrium phase boundary. An example is the XY model, where the authors showed nonanalyitcal behaviors in the LR with ground state when one quenches the system within a single phase \cite{Ddptfept}. The Hamiltonian of XY model is given by
\begin{equation}   \label{XY}
	H   =   -\sum_{j   =   1}^N \bigg[ \bigg( \frac{1   +   \delta}{2} \bigg)\sigma_j^x \sigma_{j   +   1}^x   +   \bigg( \frac{1   -   \delta}{2} \bigg)\sigma_j^y \sigma_{j   +   1}^y \bigg]   -   g \sum_{j   =   1}^N \sigma_j^z.
\end{equation}
The domains at which the system is quenched to for an arbitrary prequench Hamiltonian $   (g_i,\delta_i)   $ to achieve the single-phase DQPT is defined by the inequality\cite{Ddptfept}
\begin{equation}   \label{domain}
	\mathscr{D}(g_i,\delta_i)   =   \big\{ (g_f,\delta_f) | 2\delta_i\delta_f   <   1   -   g_ig_f   -   \sqrt{( g_i^2   -   1 )( g_f^2   -   1 )} \big\}.
\end{equation}
The XY model is diagonalizable using the same transformations in the TFIM as presented in the main text. We apply our scheme LAS in quasiparticle picture to one initial parameter pair $   (g_i,\delta_i)   =   (0,0.3)   $, where the "transition point" reckoned using Eq. (\ref{domain}) is $   g=\sqrt{0.3276}   \approx   0.573   $. We choose the final parameter pair as $   (g_f,\delta_f)   =   (0.8,0.3)   $ and observe four different contiguous excitations of \textit{k} modes:
\begin{equation}   \label{LRS}
	\begin{aligned}
		\text{(a)} \qquad \psi_n(g_i)   & =   \prod_{k'   =   k_1}^{k_{m_a}} \eta_{k'}^\dagger \eta_{-k'}^\dagger \ket{0(g_i)} \qquad m_a   =   1,2,\dots,\frac{N}{2} \\
		\text{(b)} \qquad \psi_n(g_i)   & =   \prod_{k'   =   k_{N/4   -   m_b}}^{k_{N/4   +   m_b   +   1}} \eta_{k'}^\dagger \eta_{-k'}^\dagger \ket{0(g_i)} \qquad m_b   =   0,1,2,\dots,\frac{N}{4}   -   1 \\
		\text{(c)} \qquad \psi_n(g_i)   & =   \prod_{k'   =   k_{N/2   -   m_c}}^{k_{N/2}} \eta_{k'}^\dagger \eta_{-k'}^\dagger \ket{0(g_i)} \qquad m_c   =   0,1,2,\dots,\frac{N}{2}   -   1 \\
		\text{(d)} \qquad \psi_n(g_i)   & =   \prod_{k'   =   k_{N/6}}^{k_{N/6   +   m_d}} \eta_{k'}^\dagger \eta_{-k'}^\dagger \ket{0(g_i)} \qquad m_d   =   1,2,\dots,\frac{N}{3},
	\end{aligned}
\end{equation}

\begin{figure*} [h!]
	\centering
	\hspace{-15pt}
	\includegraphics[width=18cm]{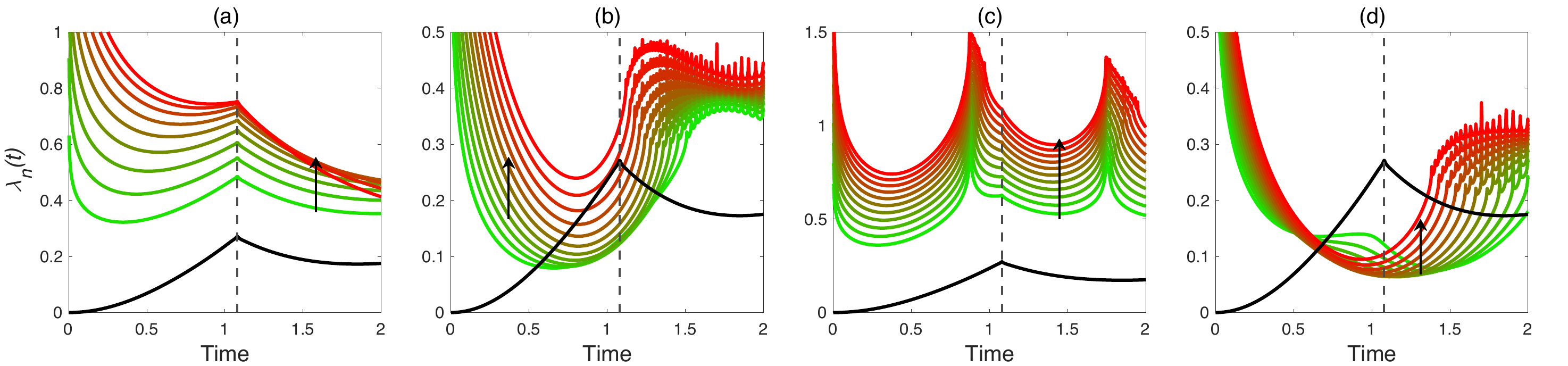}
	\caption{LRS for four distinct excited state patterns stated in Eq. (\ref{LRS}) for an $   N   =   300   $ system quenched from $   (0,0.3) \rightarrow (0.8,0.3)   $. Uparrows show the direction of increasing \textit{k} occupation. Black lines are the rate function for ground state with dashed lines indicating the critical time.}
	\label{xyrate}
\end{figure*}

\noindent where the last excitation starts from around the middle of the lower-half \textit{k} spectrum. Their respective rate functions are plotted in Fig. \ref{xyrate}. The dynamics of the system around DQPT are very similar to that being observed in the TFIM in the main text: As in Fig. \ref{xyrate}(c), there are merely excitations to very high-energy excited states. The transient excitation of the quasiparticle states are mainly in the middle of the spectrum, as shown in Fig. \ref{xyrate}(b), followed by a rapid excitation shift to around the middle range of the lower-half \textit{k} states as indicated in Fig. \ref{xyrate}(d). The only qualitative difference of the dynamics happened during DQPT between TFIM and the XY model is that the dominating excitation in the vicinity of dynamical phase transition in the former model is the entire lower-half modes, whereas the latter has a narrower range of excitation and lowest \textit{k} mode is slightly higher at around $   k_{N/6}   $. The significance of the study is that it seems to be a general case that a downshift of momentum excitation should occur when an integrable system is quenched and DQPT is triggered. There may be other minor difference for the region of excitations the system has during quench, like the negligible contribution from the lower-range \textit{k} mode excited states in XY model (Fig. \ref{xyrate}(a)) compared to that in TFIM, but the crucial dynamics around DQPT, namely the population re-distribution in the quasiparticle picture, persists in both models.

\section{Loschmidt rate for randomly selected excited state}
\label{sec:random_excited}

\begin{figure} [b!]
	\centering
	\includegraphics[width=9cm]{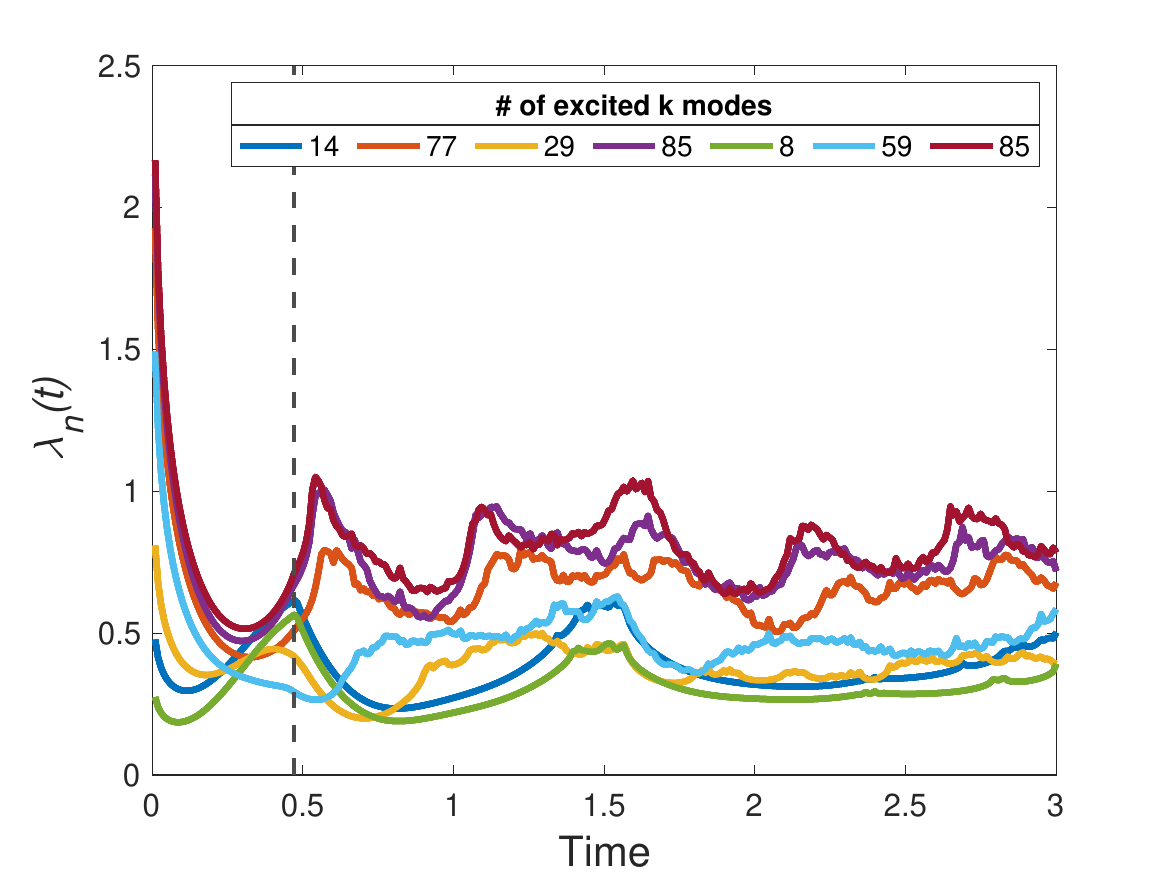}
	\caption{LRS for seven randomly excited states in an $   N   =   300   $ system. The quench is from $   g   =   0.1 \rightarrow 2   $. Dashed line indicates the first critical time. Legend shows the number of excited \textit{k} states.}
	\label{randomLRs}
\end{figure}

In the main text, the Loschmit rate function for an arbitrary excited state in the TFIM is derived in the momentum space as
\begin{equation}
	\lambda_n(t)   \sim   -\frac{1}{N} \bigg\{ \sum_{k'} \ln\bigg[ 2T_{k'}^2( 1   -   \cos( 2\varepsilon_{k'}(g_f)t ) ) \bigg]   +   \sum_{k   \neq   k'   >   0} \ln\bigg[ 1   +   T_k^4   +   2T_k^2\cos( 2\varepsilon_k(g_f)t ) \bigg] \bigg\}
\end{equation}
with $   k'   $ representing the excited \textit{k} modes. The expression contains two sums, with the former concerning all occupied momentum states and the latter concerning the empty momentum states. Since the $k$-modes are independent, one can always observe the dynamics of an arbitrarily excited state using the above equation. We show the evolution of LRs for seven different randomly excited states in Fig. \ref{randomLRs}. Note that the excited \textit{k} modes are chosen randomly in a discrete uniform distribution. Also one can hardly specify exactly which \textit{k} states are occupied due to the tremendous number of ways for a system given size $   N   $ to be excited ($   2^{N   -   1}   $ possible excited states in a parity subspace for TFIM). To that, we label those excited states by only the number of occupied \textit{k} modes, but by further inspection one can already draw some conclusions about the features of rate functions for different excited states. In particular, we divide them into several categories according to their transient growth in the LRs around the first critical time:
\begin{enumerate}
	\item Ground-state-like LRs (green and blue lines in figure \ref{randomLRs}): A main big peak similar to the nonanalytical peak of the ground-state rate function with some additional spikes of nonanalyticities along the evolution originated from the nonanalyticites in the $   \Lambda_{k'}(t)   $ defined as
	\begin{equation}   \label{LRSkp}
		\Lambda_{k'}(t)   =   \frac{1}{N}\ln\bigg[ 2T_{k'}^2( 1   -   \cos( 2\varepsilon_{k'}(g_f)t ) ) \bigg]
	\end{equation}
	with the associated critical time $   t_m(k',g_f)   =   m\pi   /   \varepsilon_{k'}(g_f),m   =   1,2,\dots   $ (for details see the original manuscript);
	\item low-range \textit{k}-excited-like LRs (cyan line in Fig. \ref{randomLRs}): Opposite to ground-state-like LRs, they have the lowest valley instead of big peak around critical time implying the most probable groups of eigenstates the system will be in during dynamical phase transitions;
	\item high-range \textit{k}-excited-like LRs (purple and brown lines in Fig. \ref{randomLRs}): Magnitudes are high in general that they merely contribute to the dynamics of DQPTs, and they have apparent nonanalytical behavior around DQPTs.
\end{enumerate}
Notice however that the LRs for some excited states may have a mixture of these categories. To name a few, the middle-range \textit{k}-mode excited state mentioned in the original manuscript and red line in Fig. \ref{randomLRs} lie in between low-range \textit{k}-excited-like and high-range \textit{k}-excited-like LRs, namely the valley moves slightly away from the critical time but still these excited states contribute to the dynamics of the quench before DQPT, and the nonanalyticity starts becoming pronounced. The yellow line in Fig. \ref{randomLRs}, on the other hand, combines the features of ground-state-like and low-range \textit{k}-excited-like LRs, where the big peak shrinks while the valley is moving closer to the critical time. To conclude, for a system that can be described by quasiparticles in momentum space, we observe several major features regarding the dynamical growth of Loschmidt rate for a general excited state. In particular, the Loschmidt rate for an arbitrarily excited state will either possess solely one of the categorized characteristics or a combination of those characteristics.
\end{widetext}

\end{document}